\documentclass[%
 reprint,
 amsmath,amssymb,
 aps,
]{revtex4-2}

\usepackage{graphicx}
\usepackage{dcolumn}
\usepackage{bm}

\begin{document}


\title{A Monte Carlo study of multiplicity fluctuations \\in proton-proton collisions at $\sqrt{s}=$~7~TeV}

\author{Valeria Zelina Reyna Ortiz}
\author{Maciej Rybczy\'nski} 
\email{maciej.rybczynski@ujk.edu.pl} 
\author{Zbigniew W\l odarczyk}%
\affiliation{Institute of Physics, Jan Kochanowski University, 25-406 Kielce, Poland}%

\date{\today}

\begin{abstract}
With large volumes of data available at LHC, it has possible to study the multiplicity distributions. 
It is interesting as well to check how well event generators can describes the properties and the behavior of 
multi-particle production processes. In this paper, we analyse the oscillatory behavior of modified combinants 
in proton-proton collisions at centre of mass energy of 7 TeV. 
\end{abstract}

\maketitle

\section{Introduction}\label{sec:introduction}

Multiplicity distributions (MDs) of charged particles produced in high-energy nuclear collisions have been extensively studied in the field of multi-particle production. The determination of multiplicity distribution is among the initial observations in new high-energy experiments, primarily because it is relatively easy to obtain such information. Furthermore, MDs provide valuable insights into the underlying production processes. Since perturbative QCD fails to fully explain the observed MDs, a range of phenomenological approaches have been employed. These approaches include dynamical methods like colored string interactions~\cite{Andersson:1983ia} and the dual-parton model~\cite{Capella:1992yb}, as well as geometrical approaches~\cite{Chen:1986ns, Hwa:1987mm} leading to the fireball model~\cite{Chou:1983xg}, 
and stochastic approaches~\cite{Dewanto:2008zz, Chew:1986qv, Chan:1990hs} that model high-energy collisions 
as branchings~\cite{Dewanto:2008zz, Chew:1986qv, Chan:1990hs} or clans~\cite{Brambilla:2006zt}.

Charged particle multiplicity distribution, $P\left(N\right)$, is usually fitted with a single negative binomial distribution (NBD)~\cite{Grosse-Oetringhaus:2009eis}:
\begin{equation}
P_{NBD}\left(N\right) = \frac{\Gamma(N+k)}{\Gamma(N+1)\Gamma(k)} p^N \left(1 - p\right)^k. \label{NBD}
\end{equation} 
NBD has two free parameters: $p$ describing probability of particle emission and parameter $k\geq 1$ influencing shape of the distribution.

Nevertheless, as the energy and the number of charged secondaries, denoted as $N$, increase, the negative binomial distribution tends to deviate from the observed data for large values of $N$, as discussed in~\cite{Wilk:2016dcn}. In these cases, alternative approaches are adopted, including combinations of two~\cite{Ghosh:2012xh, Giovannini:2003ft}, three~\cite{Zborovsky:2013tla}, or multi-component NBDs~\cite{Dremin:2004ts}, or even different forms of $P\left(N\right)$ 
distributions~\cite{Andersson:1983ia, Grosse-Oetringhaus:2009eis, Dremin:2000ep, Chekanov:1996ah, Hoang:1987tt}. However, it should be noted that such adjustments primarily improve the agreement for large $N$, while the ratio $R = \text{data/fit}$ exhibits significant deviations from unity at small $N$ across all fitting scenarios~\cite{Wilk:2016dcn, Wilk:2018kvg}.

Such a observation suggests that there is additional information in the measured multiplicity distribution that is not covered by the following recurrence relation:
\begin{equation}
\left(N+1\right)P\left(N+1\right) = \gamma\left(N\right)P\left(N\right),\quad \gamma\left(N\right) = \alpha + \beta N.\label{rr1}
\end{equation}
Three commonly encountered forms of $P(N)$ resulting from the recurrence relation (\ref{rr1}) are as follows: the binomial distribution, where $\alpha = Kp/(1-p)$ and $\beta = -\alpha/K$; the Poisson distribution, where $\alpha = \lambda$ and $\beta = 0$; and the negative binomial distribution, where $\alpha = kp$ and $\beta = \alpha/k$. Here, the parameter $p$ again represents the probability of particle emission. In our previous work~\cite{Wilk:2016dcn}, we introduced a more generalized form of the recurrence relation that is applicable in counting statistics when considering multiplication effects in point processes~\cite{ST}. Unlike Eq.~(\ref{rr1}), this new relation connects all multiplicities using coefficients $C_{j}$, which determine the corresponding $P(N)$ in the following manner:
\begin{equation}
\left(N + 1\right)P\left(N + 1\right) = \langle N\rangle \sum^{N}_{j=0} C_j P\left(N - j\right). \label{rr2}
\end{equation}
The modified combinants, $C_{j}$, can be obtained from experimental data
\begin{equation}
\langle N\rangle C_j = (j+1)\left[ \frac{P(j+1)}{P(0)} \right] - \langle N\rangle \sum^{j-1}_{i=0}C_i \left[ \frac{P(j-i)}{P(0)} \right] \label{rCj}
\end{equation}
and exhibit a pronounced oscillatory pattern. This behavior is not only observed in proton-proton collisions, as discussed in previous works such as~\cite{Wilk:2016dcn, Rybczynski:2018bwk, Rybczynski:2019dwa, Zborovsky:2018vyh}, but has also been recently demonstrated in $e^{+}e^{-}$ annihilation processes, as shown in~\cite{Ang:2018zjy}. These oscillations suggest the presence of additional information regarding the multi-particle production process that remains undisclosed. The periodic nature of the oscillations observed in the modified combinants derived from experimental data is particularly indicative in this regard.

Nevertheless the probability that such oscillations are statistically insignificant is very small (${\sim}10^{-16}$, see Ref.~\cite{Rybczynski:2018bwk} for more details) the sensitivity to experimental procedures are still under debate. 

The aim of this paper is to show that the observed oscillations have a physical origin and are not the result of experimental procedures. We focus on the analysis of the Monte Carlo simulated events and comparison with existing experimental data. The paper is organized as follows. In Sec.~\ref{sec:models} we discuss the methodology of event generation and analysis of model data. In Sec.~\ref{sec:results} we provide a concise description of the results we obtained for proton-proton interactions. Finally, in Sec.~\ref{sec:discussion} we made several comments referring to the oscillatory behavior of the higher-order moments of multiplicity distributions observed both in experimental data and the models.

\section{Event generation and analysis methodology} \label{sec:models}

PYTHIA~\cite{Bierlich:2022pfr} is a widely used Monte Carlo event generator program designed to generate events in high-energy physics. It serves as a tool for simulating collisions at high-energies involving elementary particles like $e^+$, $e^-$, protons, anti-protons, as well as heavy-ions, and various combinations thereof. The program encompasses a wide range of physics aspects, such as total and partial cross sections, interactions at both hard and soft scales, parton distributions, initial- and final-state parton showers, matching and merging of matrix elements with showers, multi-parton interactions, as well as processes related to hadronization/fragmentation and particle decays. 

The Energy Parton Off-shell Splitting (EPOS) transport model~\cite{Pierog:2013ria} is a Monte Carlo event generator program designed for simulation high-energy particle collisions. It provides a framework to study various aspects of particle interactions and the resulting hadron production in both nucleus-nucleus and proton-proton collisions. EPOS considers each nucleus-nucleus or proton-proton collision as a collection of many elementary collisions happening simultaneously. These collisions involve the exchange of ''parton ladders``, which represent the evolution of partons from the projectile and target sides towards the central region (small x). The evolution of partons in EPOS is governed by an evolution equation, typically based 
on the Dokshitzer-Gribov-Lipatov-Altarelli-Parisi (DGLAP) formalism. The intermediate gluons in the parton ladder are treated as kink singularities within the framework of relativistic strings. These strings represent flux tubes that connect the interacting partons. These flux tubes eventually decays by producing 
quark-antiquark pairs, giving rise to fragments that are identified as hadrons~\cite{Motornenko:2017klp}.

The Ultra-relativistic Quantum Molecular Dynamics (UrQMD) model, as outlined in~\cite{Bass:1998ca, Bleicher:1999xi}, is a microscopic transport model that employs the covariant propagation of hadrons along classical trajectories, coupled with stochastic binary scatterings, color string formation, and resonance decay. It provides a comprehensive framework to study the dynamics and interactions of particles across a wide energy range. It operates as a Monte Carlo solution to a complex system of coupled partial integro-differential equations, which describe the time evolution of phase space densities for various particle species. In the UrQMD model, baryon-baryon collisions at lower energies consider the exchange of electric and baryonic charge, strangeness, and four-momentum in the t-channel. On the other hand, meson-baryon and meson-meson interactions are treated through the formation and subsequent decay of resonances, following the s-channel reaction mechanism. At higher energies, a vast array of particle species can be generated, and the model accounts for subsequent re-scatterings. The UrQMD model allows generate all types of particles in hadron-hadron collisions and enables their further interaction with one another~\cite{Motornenko:2017klp}. For the analysis described in this paper, we used UrQMD set to LHC mode. It means that no hydrodynamics functions were activated. This compilation mode of the UrQMD model, in short words, is prepared to do calculations over high values of multiplicity. No essential changes to the model parameters were introduced.

In this study we have used PYTHIA 8.308~\cite{Bierlich:2022pfr}, EPOS LHC~\cite{Pierog:2013ria} and UrQMD 3.4 set to LHC mode~\cite{Bass:1998ca} to generate proton-proton interactions at $\sqrt{s}=7$~TeV in accordance to data on charged particles multiplicity distributions obtained by the ALICE experiment at CERN LHC~\cite{ALICE:2017pcy}. In PYTHIA simulation we have implemented the inelastic component of the total cross-section for soft-QCD processes with the parameter \emph{SoftQCD:inelastic=on}. The remaining set of PYTHIA parameters we left with its default values. In the case of EPOS LHC and UrQMD we used default values of the parameters. To match with the experimental conditions, charged particle multiplicities have been chosen in the trigger conditions and acceptance of the ALICE detector, defined in~\cite{ALICE:2017pcy}. Namely, the generated events of collisions (EOCs) were divided into two classes: inelastic (\emph{INEL}) class and non-single diffractive (\emph{NSD}) class. The generated EOC belongs to INEL class if there is at least one charged particle in either the $-3.7<\eta<-1.7$, $|\eta|<2.98$, or $2.8<\eta<5.1$ pseudorapidity interval corresponding to the acceptances of the V0-C, SPD, and V0-A ALICE sub-detectors, respectively. The NSD class requires charged particles to the detected in both $-3.7<\eta<-1.7$ and $2.8<\eta<5.1$ pseudorapidity intervals~\cite{ALICE:2017pcy}.

\section{Results}\label{sec:results}

\begin{figure}
\begin{center} 
\includegraphics [width=\linewidth]{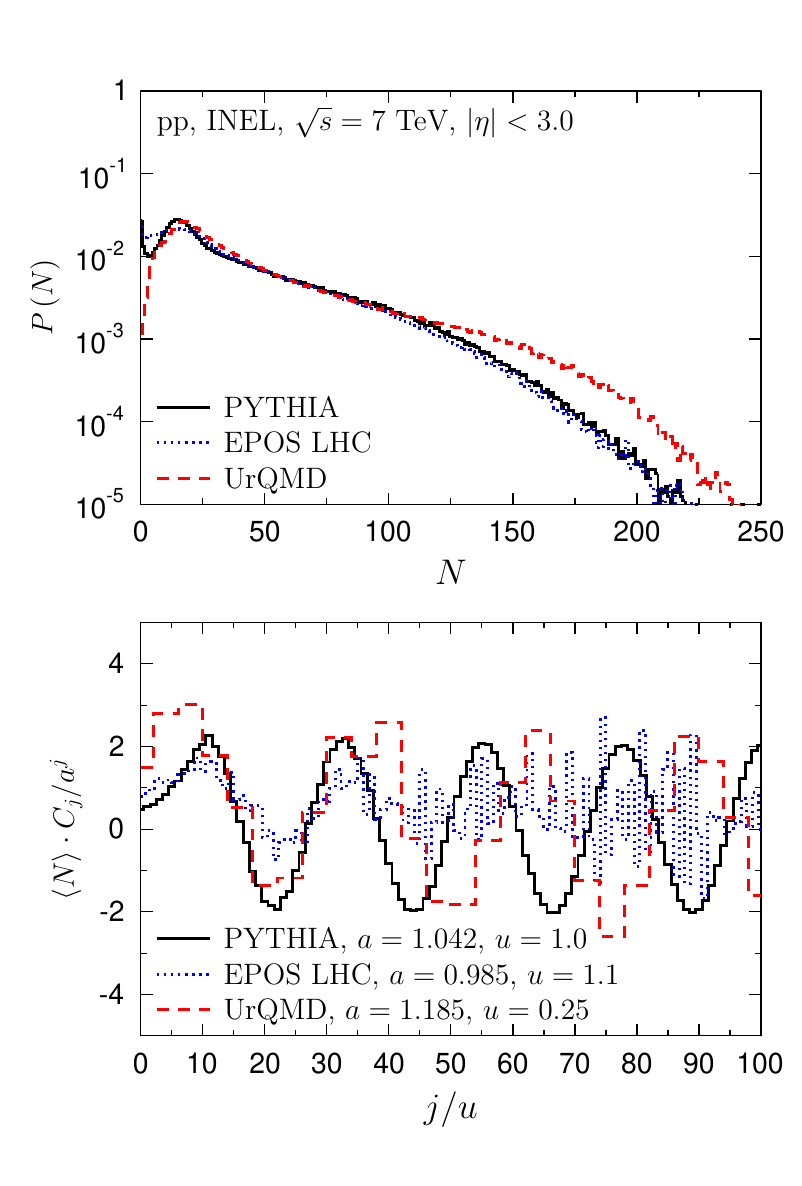}
\end{center}
\caption{(upper panel) Multiplicity distributions $P\left(N\right)$ of charged particles generated in proton-proton interactions at $\sqrt{s}=7$~TeV. (lower panel) The corresponding modified combinants $C_{j}$ emerging from them. PYTHIA 8 ~\cite{Bierlich:2022pfr} with \emph{SoftQCD:inelastic} processes (solid lines), EPOS LHC~\cite{Pierog:2013ria} (dotted lines) and UrQMD 3.4 set to LHC mode~\cite{Bass:1998ca} (dashed lines). For all models the applied kinematic cuts as described in the ALICE paper~\cite{ALICE:2017pcy} ({\it INEL} class).\label{fig:inel}}
\end{figure}

\begin{figure}
\begin{center} 
\includegraphics [width=\linewidth] {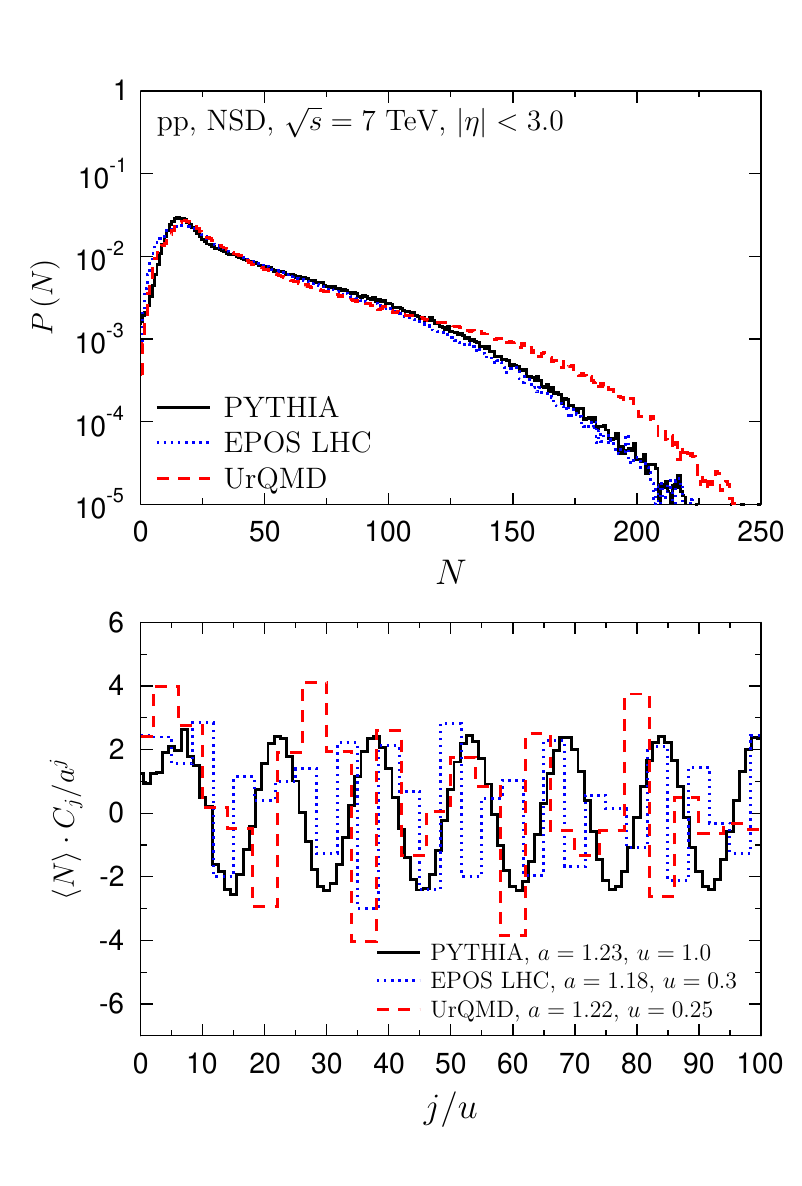}
\end{center}
\caption{ Same as Fig.~\ref{fig:inel}, but for {\it NSD} class.\label{fig:nsd}}
\end{figure}

\begin{figure}
\begin{center} 
\includegraphics [width=\linewidth] {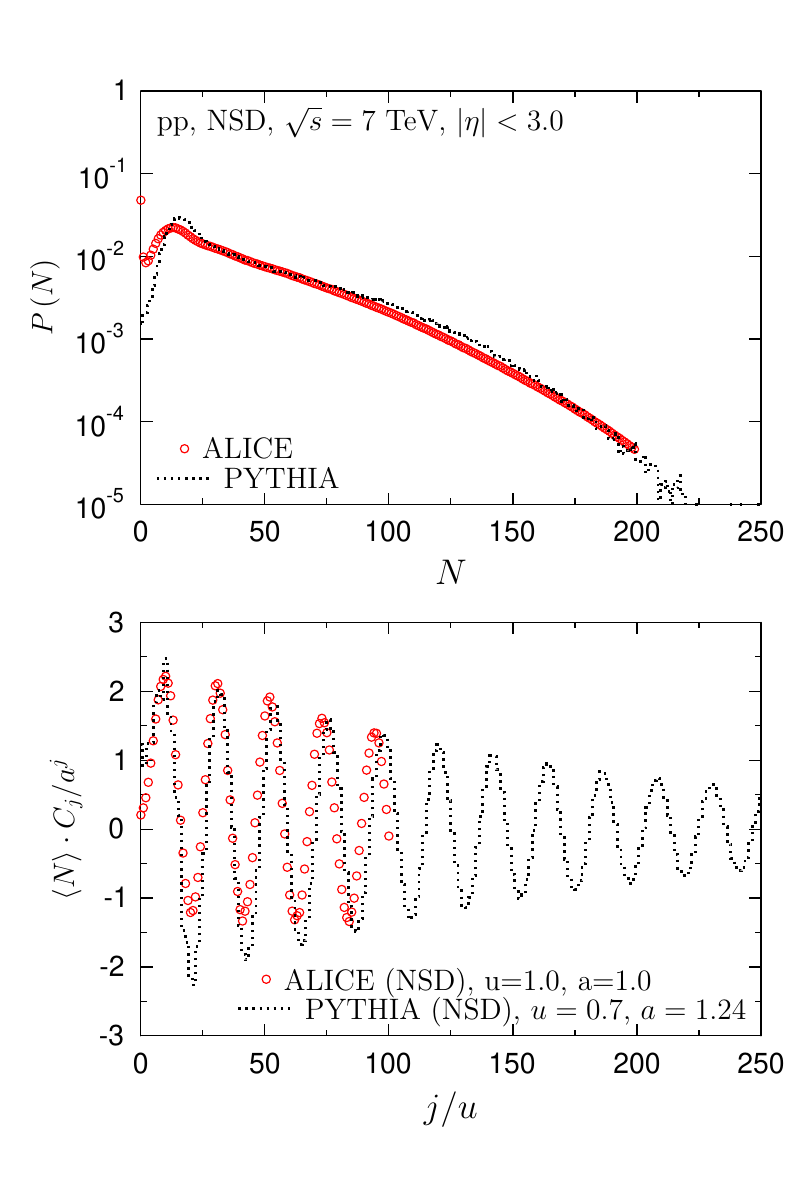}
\end{center}
\caption{(upper panel) Multiplicity distribution $P\left(N\right)$ of charged particles produced in proton-proton {\it non-single diffractive} interactions at $\sqrt{s}=7$~TeV as measured by ALICE experiment~\cite{ALICE:2017pcy} (NSD class). (lower panel) The corresponding modified combinants $C_{j}$ emerging from them. PYTHIA 8~\cite{Bierlich:2022pfr} (dotted lines) with \emph{SoftQCD:inelastic} processes and all kinematic cuts as described in the ALICE paper.\label{fig:fignsd_like_in_exp_pythia}}
\end{figure}
\begin{figure}
\begin{center} 
\includegraphics [width=\linewidth]{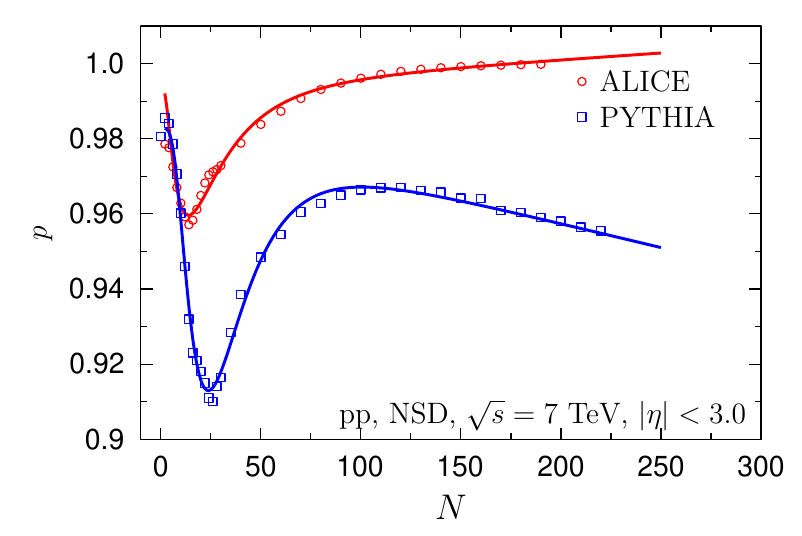}
\end{center}
\caption{Values of NBD $p$ parameter chosen to obtain NBD fits. See text for details. \label{fig:fig_nbd}}
\end{figure}
\begin{figure}
\begin{center} 
\includegraphics [width=\linewidth]{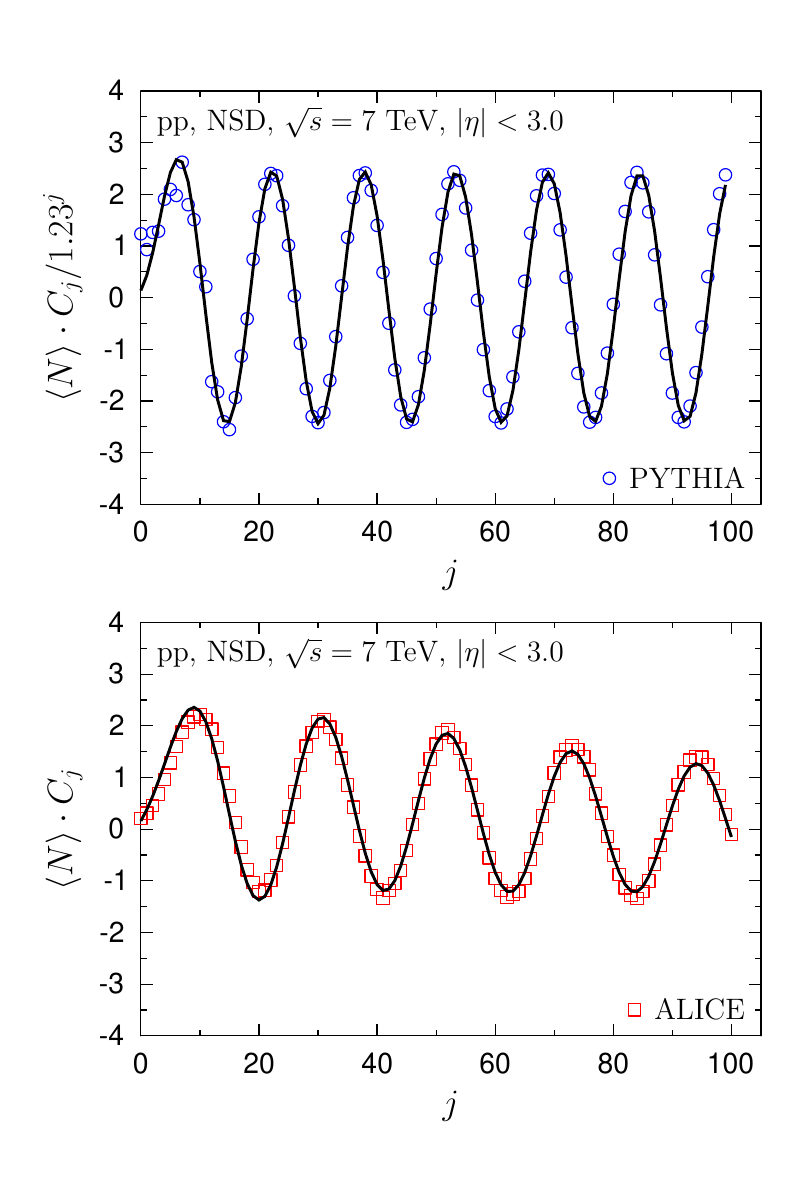}
\end{center}
\caption{(upper panel) Modified combinants $C_{j}$ calculated from multiplicity distribution of charged particles generated in proton-proton interactions at $\sqrt{s}=7$~TeV using PYTHIA 8 ~\cite{Bierlich:2022pfr} with \emph{SoftQCD:inelastic} processes and kinematic cuts as described in the ALICE paper~\cite{ALICE:2017pcy} ({\it NSD} class). (lower panel) The corresponding modified combinants calculated from ALICE data~\cite{ALICE:2017pcy} (NSD class). Solid lines in both panels show fits obtained with parameters listed in Table~\ref{tab-2}.\label{fig:fig_tab2}}
\end{figure}

Multiplicity distributions $P\left(N\right)$ of charged particles in simulated events and the modified combinants $C_j$ that results from them are shown in Fig.~\ref{fig:inel} for \emph{INEL} class and in Fig.~\ref{fig:nsd} for \emph{NSD} class. Modified combinants in all models exhibit oscillating behavior, however their amplitudes and periods of oscillations are different. Growth of amplitudes with rank $j$ can be described as $\langle N\rangle C_{j} \sim a^{j}$ with $a=$ 1.042, 0.985 and 1.185 (\emph{INEL} class), and $a=$ 1.23, 1.18 and 1.22 (\emph{NSD} class) for PYTHIA, EPOS and UrQMD models respectively. Periods of oscillations (\emph{INEL} class) are 22, 28 and 7 for PYTHIA, EPOS and UrQMD models respectively. For \emph{NSD} class, periods of oscillations are smaller and equal to 16, 4 and 3, respectively.
Remarkable oscillations of modified combinants $C_{j}$ and multiplicity distribution $P(N)$ given by PYTHIA model are compared with experimental data in Fig.~\ref{fig:fignsd_like_in_exp_pythia}. The model and ALICE data show substantial discrepancies at small multiplicities, the experimental results for $P(N)$ cannot be described exactly. In particular, for the void probability $P\left(0\right)$, we observe large difference between ALICE data and PYTHIA prediction. Since
\begin{equation}
 P\left(0\right) = \exp\left(-\sum_{j=0}^{\infty} \frac{\langle N\rangle C_{j}}{j+1}\right)
\label{eq:void}
\end{equation}
this find reflection in behavior of modified combinants. Comparing with ALICE data, in PYTHIA model the period of oscillations is 1.43 times larger and the ratio of amplitudes (model/data) increase as $1.24^j$. 

\begin{table}
\centering
\caption{Parameters $A$, $B$, $C$, $D$ and $E$ of the function $p\left(N)\right)$ given by Eq.~(\ref{eq:pn}), used to fit the data in Fig. ~\ref{fig:fig_nbd}.}
\label{tab-1}
\vspace{2mm}
\begin{tabular}{cccccc}
\hline
& $A$ & $B$  & $C$  & $D$ & $E$ \\
\hline
ALICE~~ & $0.994$ & $-3.6\cdot 10^{-5}$ & $0.68$ & $27$ & $1.15$ \\
PYTHIA~ & $0.983$ &~ $1.3 \cdot 10^{-4}$ & $1.9$ & $33$ & $0.83$ \\
\hline
\end{tabular}
\end{table}

The most commonly used form of $P\left(N\right)$, the NBD form given by Eq.~(\ref{NBD}) does not describe experimental data nor the multiplicity distributions given by models. To describe $P\left(N\right)$ using NBD, the negative binomial distribution parameter $p$ must depend on $N$~\cite{Wilk:2016dcn}. The probability of particle emission, $p$, which is constant in the standard NBD, is dependent on the multiplicity $N$ in the way presented in Fig.~\ref{fig:fig_nbd}. Both in experimental data and model the non-monotonic form of $p$ is clearly visible and can be described by
\begin{equation}
\begin{split}
p\left(N\right) = & A\left(1-B\cdot N\right) \cdot \\ 
& \left(1-C/N\exp\left(-\left(\ln\left(N/D\right)/E\right)^{2}\right)\right)
\end{split}
\label{eq:pn}
\end{equation}
with parameters listed in Table~\ref{tab-1}. Nevertheless the behaviors are roughly similar, we observe significant difference in position $N=D\exp{(-E^2/2)}$ of minimum of $p$. The probability of particle emission $p=min$ for multiplicity $N \simeq 23$ in PYTHIA model and $N \simeq 14$ for ALICE experimental data. 

\begin{table*}
\caption{Parameters $w_i$, $p_i$, $K_i$, $k_i$ and $m_i$ of the $2$-component $P(N)$ given by Eq.~(\ref{eq:pn2}), 
used to fit the ALICE data and corresponding PYTHIA simulation, $\sqrt{s}=7$~TeV and $|\eta|< 3$.}
\centering
\label{tab-2}
\vspace{2mm}
\begin{tabular}{*{11}{p{0.08\linewidth}}}
\hline
& $w_1$ & $p_1$  & $K_1$  & $k_1$ & $m_1$  & $w_2$  & $p_2$  &$K_2$& $k_2$ & $m_2$~  \\
\hline
 ALICE~~ &~ $0.24$ & $0.90$ & $3$ & $2.80$ & $5.75$ & $0.76$ & $0.645$ & $3$ & $1.34$ & $23.5$~ \\
 PYTHIA~ &~ $0.52$ & $0.97$ & $3$ & $2.85$ & $6.75$ & $0.48$ & $0.790$ & $3$ & $2.30$ & $27.5$~ \\
\hline
\end{tabular}
\end{table*}

Multiplicity distributions measured by ALICE can be successfully described by a two-component compound binomial distribution 
\begin {equation}
P(N)=\sum_{i=1}^2 w_i h(N;p_i,K_i,k_i,m_i),
\label{eq:pn2}
\end {equation}
where $h(N)$ is the compound binomial distribution (BD + NBD) given by the generating function
\begin {equation}
H(z)=\left [ p \left ( \frac{k}{k-m(z-1)}\right)^k +1-p \right]^K.
\label{eq:gf}
\end{equation}
Comparison with PYTHIA simulation is shown in Fig.~\ref{fig:fig_tab2} for parameters given in Table~\ref{tab-2}.


\section{Discussion}\label{sec:discussion}

It is worth pointing out that modified combinants evaluated from models exhibit oscillatory behavior, though the oscillation period differs from experimental data. Modified combinants $C_{j}$ can be expressed by the generating function:
\begin{equation}
 G\left(z\right) = \sum_{N=0}^{\infty} P\left(N\right) z^{N}
\label{eq:gen_fun}
\end{equation}
of the count probability $P\left(N\right)$ as:
\begin{equation}
\langle N\rangle C_{j} = \frac{1}{j!}\frac{d^{j+1}\ln G\left(z\right)}{dz^{j+1}}\Bigg|_{z=0}.
\label{eq:mod_com}
\end{equation}
The generating function can be shown to be a sum over the \emph{``averaged''} connected correlation function $\overline{g_{(n)}}$ of all orders, $\left(n\right)$:
\begin{equation}
 \ln G\left(z\right) = \sum_{n=1}^{\infty} \frac{\left(z-1\right)^{n}}{n!} m^{n}\overline{g_{(n)}},
\label{eq:gen_funjj}
\end{equation}
where $m$ is the number of particles in a cell of the phase-space volume~\cite{White:1979kp}. The modified combinants $C_{j}$ can be expressed as an infinite series of the $\overline{g_{(n)}}$:
\begin{equation}
\langle N\rangle C_{j} = \sum_{n=j+1}^{\infty} \left(-1\right)^{n-j-1}\frac{m^{n}}{j!\left(n-j-1\right)!}\overline{g_{(n)}}.
\label{eq:mod_com2}
\end{equation}
Note that the correlation functions $\overline{g_{(n)}}$ are associated with widely used cumulant factorial moments (see Appendix for more details).

\begin{figure}
\begin{center} 
\includegraphics [width=\linewidth]{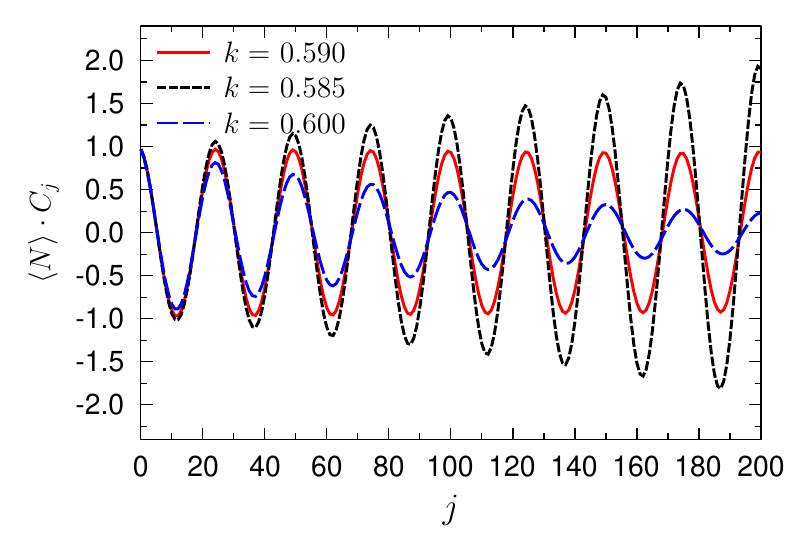}
\end{center}
\caption{Modified combinants $C_{j}$ calculated for $m=4$ and different values of $k$ parameter.}
\label{fig:fig_amp}
\end{figure}
\begin{figure}
\begin{center} 
\includegraphics [width=\linewidth]{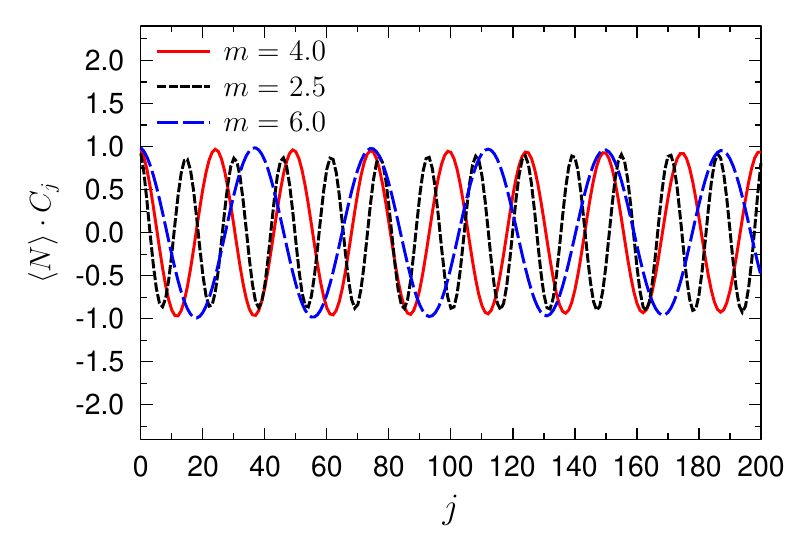}
\end{center}
\caption{Modified combinants $C_{j}$ calculated for different values of $m$ parameter. The values of $k$ parameter were adjusted in order to obtain the same amplitude of oscillations.}
\label{fig:fig_freq}
\end{figure}

Higher-order correlations, characterized by an $n-$body correlation function $g_{(n)}$, are of general interest and have been investigated in many fields of physics, including astronomy, particle physics and quantum optics. In 1963, Glauber predicted that the maximal value of the same-point normalized $n-$body correlation function $g_{(n)}$ calculated for thermal light is directly related to the order of the function by a simple relationship $n!$~\cite{Glauber:1963fi}. This $n!$ dependence is a consequence of Wick's theorem~\cite{Wick:1950ee}, which enables higher-order correlations to be expressed using products of one-body correlation functions. The applicability of Wick's theorem is not limited to correlation functions for light, it has also been applied in many other fields; for example, it is commonly used in radio-astronomy, nuclear physics, and generally in quantum field theory. The validity of Wick's theorem has been firstly demonstrated with thermal photons, and recently proved to higher-order correlations for massive particles~\cite{Dall:2013}.

For correlation function
\begin{equation}
\overline{g_{(n)}} = \left(n-1\right)!k^{-n+1}
\label{eq:corr_fun}
\end{equation}
with real positive parameter $k$ we have:
\begin{equation}
\langle N\rangle C_{j} = k\left(m/\left(m+k\right)\right)^{j+1}
\label{eq:mod_com3}
\end{equation}
as for NBD, where $m$ is the average multiplicity and the two-body correlation function determines the value of NBD shape parameter, $1/k=g_{(2)}$.

To assure oscillating behavior of modified combinants we choose correlation function in the form:
\begin{equation}
\overline{g_{(n)}} = \left(n-1\right)!\cos\left(n/k\right),
\label{eq:corr_fun2}
\end{equation}
which leads to to the following formula for modified combinants:
\begin{equation}
\langle N\rangle C_{j} = \frac{1}{2}m^{j+1}\left[\left(e^{\imath/k}+m\right)^{-j-1}+\left(e^{-\imath/k}+m\right)^{-j-1}\right]
\label{eq:mod_com4}
\end{equation}
with $\imath$ denoting imaginary unit.

It is remarkable that $k$ parameter only affects oscillation amplitude, see Fig.~\ref{fig:fig_amp}. The period of oscillations ($\sim 2\pi m$) is determined by the value of $m$ parameter, see Fig.~\ref{fig:fig_freq}.

The average number of particles in a cell given by the value of $m$ parameter is not derived from any first principle, although some suggestions have been made to equate it with the average number of partons in the QCD cascade. Considering hadrons as a dense system of partons we expect that $m\sim Q_{S}^{2}$, where $Q_{S}$ denotes the saturation scale~\cite{Gotsman:2020bjc, Kharzeev:2017qzs}.

\begin{figure}
\begin{center} 
\includegraphics [width=\linewidth]{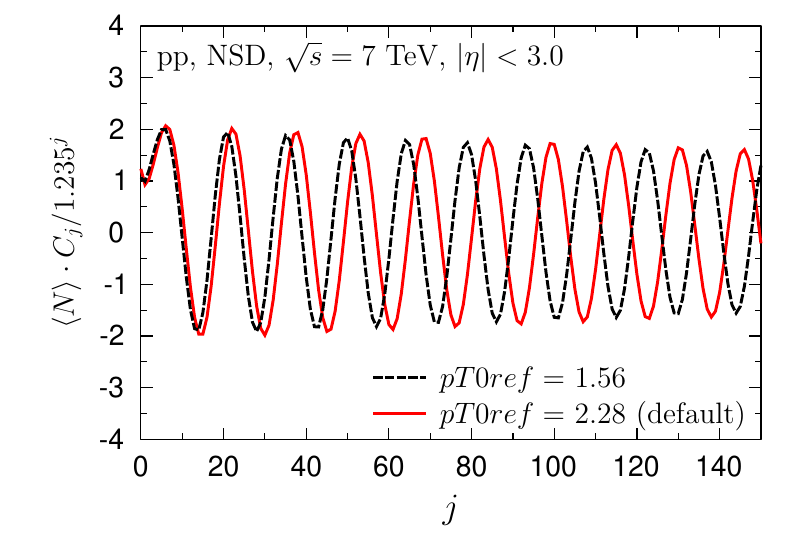}
\end{center}
\caption{Modified combinants $C_{j}$ calculated from multiplicity distribution of charged particles generated in proton-proton interactions at $\sqrt{s}=7$~TeV using PYTHIA 8 ~\cite{Bierlich:2022pfr} with \emph{SoftQCD:inelastic} processes and kinematic cuts as described in the ALICE paper~\cite{ALICE:2017pcy} ({\it NSD} class). Simulation performed for two values of \emph{pT0Ref} parameter. Using red solid line we show the results for \emph{pT0Ref}=2.28, which is default PYTHIA value. Black dashed line show the result obtained for \emph{pT0Ref}=1.56.}
\label{fig:fig_pt0ref}
\end{figure}

In the PYTHIA model, the saturation scale $Q_{S}$ can be connected with \emph{$p_{T0}$} parameter~\cite{Bierlich:2022pfr}.  The period of oscillations depends on \emph{$p_{T0}$}, but $\sim p_{T0}^{1/3}$ dependence is very weak. By setting \emph{MultiPartonInteraction:pT0Ref=1.56} instead of its default value $2.28$ we found the period of oscillations decreasing by $15\%$ (see Fig.~\ref{fig:fig_pt0ref}), but a more thorough re-tune is necessary in order to simultaneously obtain the correct multiplicity distribution.

Our results can, hopefully, lead to wider theoretical investigations and provide a  better understanding of multi-particle production processes in hadronic collisions.\\

\appendix*
\section{Relationship of correlation functions with cumulants}
\label{appendix}
Usually information contained in $P\left(N\right)$ is obtained by examining their corresponding cumulant factorial moments~\cite{Bartel}
\begin{equation}
K_q=F_q-\sum_{i=1}^{q-1} \binom{q-1}{i-1}K_{q-i}F_i,
\label{eq:kq}
\end{equation}
where
\begin{equation}
F_q=\sum_{N=q}^{\infty}N(N-1)(N-2)...(N-q+1)P(N)
\label{eq:fg}
\end{equation}
are the factorial moments. Modified combinants $C_j$ can be expressed as an infinite series of the $K_q$~\cite{Rybczynski:2018bwk}
\begin{equation}
\langle N\rangle C_j=\frac{1}{j!} \sum_{p=0}^\infty \frac{(-1)^p}{p!}K_{p+j}.
\label{eq:cj}
\end{equation}
When comparing Eq.~(\ref{eq:cj}) with Eq.~(\ref{eq:mod_com2}) we have
\begin{equation}
K_q=m^{q+1} \overline{g_{(q+1)}}.
\label{eq:kqgq}
\end{equation}


Note that the moments $K_q$ require knowledge of all $P\left(N\right)$ while, according to Eq.~(\ref{rCj}), calculation of $C_j$ (and corresponding correlation function $\overline{g_{(n)}}$) requires only a finite number of probabilities $P\left(N<j\right)$ which may be advantageous in applications.
 
\section*{Acknowledgements}
This research was supported by the Polish National Science Centre (NCN) Grant No. 2020/39/O/ST2/00277. In preparation of this work we used the resources of the Center for Computation and Computational Modeling of the Faculty of Exact and Natural Sciences of the Jan Kochanowski University of Kielce. 


\begin{thebibliography}{99}

\bibitem{Andersson:1983ia}
B.~Andersson, G.~Gustafson, G.~Ingelman and T.~Sjostrand,
Phys. Rept. \textbf{97}, 31-145 (1983)
doi:10.1016/0370-1573(83)90080-7

\bibitem{Capella:1992yb}
A.~Capella, U.~Sukhatme, C.~I.~Tan and J.~Tran Thanh Van,
Phys. Rept. \textbf{236}, 225-329 (1994)
doi:10.1016/0370-1573(94)90064-7

\bibitem{Chen:1986ns}
W.~R.~Chen and R.~C.~Hwa,
Phys. Rev. D \textbf{36}, 760 (1987)
doi:10.1103/PhysRevD.36.760

\bibitem{Hwa:1987mm}
R.~C.~Hwa,
Phys. Rev. D \textbf{37}, 1830 (1988)
doi:10.1103/PhysRevD.37.1830

\bibitem{Chou:1983xg}
K.~c.~Chou, L.~s.~Liu and T.~c.~Meng,
Phys. Rev. D \textbf{28}, 1080 (1983)
doi:10.1103/PhysRevD.28.1080

\bibitem{Dewanto:2008zz}
A.~Dewanto, A.~H.~Chan, C.~H.~Oh, R.~Chen and K.~Sitaram,
Eur. Phys. J. C \textbf{57}, 515-523 (2008)
doi:10.1140/epjc/s10052-008-0750-z

\bibitem{Chew:1986qv}
C.~K.~Chew, D.~Kiang and H.~Zhou,
Phys. Lett. B \textbf{186}, 411-415 (1987)
doi:10.1016/0370-2693(87)90318-2

\bibitem{Chan:1990hs}
A.~H.~Chan and C.~K.~Chew,
Phys. Rev. D \textbf{41}, 851-862 (1990)
doi:10.1103/PhysRevD.41.851

\bibitem{Brambilla:2006zt}
M.~Brambilla, A.~Giovannini and R.~Ugoccioni,
Physica A \textbf{387}, 1110-1122 (2008)
doi:10.1016/j.physa.2007.10.047
[arXiv:hep-ph/0605269 [hep-ph]].

\bibitem{Grosse-Oetringhaus:2009eis}
J.~F.~Grosse-Oetringhaus and K.~Reygers,
J. Phys. G \textbf{37}, 083001 (2010)
doi:10.1088/0954-3899/37/8/083001
[arXiv:0912.0023 [hep-ex]].

\bibitem{Wilk:2016dcn}
G.~Wilk and Z.~W\l{}odarczyk,
J. Phys. G \textbf{44}, no.1, 015002 (2017)
doi:10.1088/0954-3899/44/1/015002
[arXiv:1601.03883 [hep-ph]].

\bibitem{Ghosh:2012xh}
P.~Ghosh,
Phys. Rev. D \textbf{85}, 054017 (2012)
doi:10.1103/PhysRevD.85.054017
[arXiv:1202.4221 [hep-ph]].

\bibitem{Giovannini:2003ft}
A.~Giovannini and R.~Ugoccioni,
Phys. Rev. D \textbf{68}, 034009 (2003)
doi:10.1103/PhysRevD.68.034009
[arXiv:hep-ph/0304128 [hep-ph]].

\bibitem{Zborovsky:2013tla}
I.~Zborovsk\'y,
J. Phys. G \textbf{40}, 055005 (2013)
doi:10.1088/0954-3899/40/5/055005
[arXiv:1303.7388 [hep-ph]].

\bibitem{Dremin:2004ts}
I.~M.~Dremin and V.~A.~Nechitailo,
Phys. Rev. D \textbf{70}, 034005 (2004)
doi:10.1103/PhysRevD.70.034005
[arXiv:hep-ph/0402286 [hep-ph]].

\bibitem{Dremin:2000ep}
I.~M.~Dremin and J.~W.~Gary,
Phys. Rept. \textbf{349}, 301-393 (2001)
doi:10.1016/S0370-1573(00)00117-4
[arXiv:hep-ph/0004215 [hep-ph]].

\bibitem{Chekanov:1996ah}
S.~V.~Chekanov and V.~I.~Kuvshinov,
J. Phys. G \textbf{22}, 601-610 (1996)
doi:10.1088/0954-3899/22/5/007
[arXiv:hep-ph/9606202 [hep-ph]].

\bibitem{Hoang:1987tt}
T.~F.~Hoang and B.~Cork,
Z. Phys. C \textbf{36}, 323 (1987)
doi:10.1007/BF01579149

\bibitem{Wilk:2018kvg}
G.~Wilk and Z.~W\l{}odarczyk,
Int. J. Mod. Phys. A \textbf{33}, no.10, 1830008 (2018)
doi:10.1142/S0217751X18300089
[arXiv:1803.07832 [hep-ph]].

\bibitem{ST} 
B.E.A.~Saleh and M.K.~Teich, 
Proc. IEEE {\bf 70}, 229 (1982).

\bibitem{Rybczynski:2018bwk}
M.~Rybczynski, G.~Wilk and Z.~W\l{}odarczyk,
Phys. Rev. D \textbf{99}, no.9, 094045 (2019)
doi:10.1103/PhysRevD.99.094045
[arXiv:1811.07197 [hep-ph]].

\bibitem{Rybczynski:2019dwa}
M.~Rybczy\'nski, G.~Wilk and Z.~W\l{}odarczyk,
Ukr. J. Phys. \textbf{64}, no.8, 738-744 (2019)
doi:10.15407/ujpe64.8.738
[arXiv:1906.11531 [hep-ph]].

\bibitem{Zborovsky:2018vyh}
I.~Zborovsk\'y,
Eur. Phys. J. C \textbf{78}, no.10, 816 (2018)
doi:10.1140/epjc/s10052-018-6287-x
[arXiv:1811.11230 [hep-ph]].

\bibitem{Ang:2018zjy}
H.~W.~Ang, M.~Ghaffar, A.~H.~Chan, M.~Rybczy\'nski, Z.~W\l{}odarczyk and G.~Wilk,
Mod. Phys. Lett. A \textbf{34}, no.39, 1950324 (2019)
doi:10.1142/S0217732319503243
[arXiv:1812.08840 [hep-ph]].

\bibitem{Bierlich:2022pfr}
C.~Bierlich, S.~Chakraborty, N.~Desai, L.~Gellersen, I.~Helenius, P.~Ilten, L.~L\"onnblad, S.~Mrenna, S.~Prestel and C.~T.~Preuss, \textit{et al.}
doi:10.21468/SciPostPhysCodeb.8
[arXiv:2203.11601 [hep-ph]].

\bibitem{Pierog:2013ria}
T.~Pierog, I.~Karpenko, J.~M.~Katzy, E.~Yatsenko and K.~Werner,
Phys. Rev. C \textbf{92}, no.3, 034906 (2015)
doi:10.1103/PhysRevC.92.034906
[arXiv:1306.0121 [hep-ph]].

\bibitem{Motornenko:2017klp}
A.~Motornenko, K.~Grebieszkow, E.~Bratkovskaya, M.~I.~Gorenstein, M.~Bleicher and K.~Werner,
J. Phys. G \textbf{45}, no.11, 115104 (2018)
doi:10.1088/1361-6471/aae149
[arXiv:1711.07789 [nucl-th]].

\bibitem{Bass:1998ca}
S.~A.~Bass, M.~Belkacem, M.~Bleicher, M.~Brandstetter, L.~Bravina, C.~Ernst, L.~Gerland, M.~Hofmann, S.~Hofmann and J.~Konopka, \textit{et al.}
Prog. Part. Nucl. Phys. \textbf{41}, 255-369 (1998)
doi:10.1016/S0146-6410(98)00058-1
[arXiv:nucl-th/9803035 [nucl-th]].

\bibitem{Bleicher:1999xi}
M.~Bleicher, E.~Zabrodin, C.~Spieles, S.~A.~Bass, C.~Ernst, S.~Soff, L.~Bravina, M.~Belkacem, H.~Weber and H.~Stoecker, \textit{et al.}
J. Phys. G \textbf{25}, 1859-1896 (1999)
doi:10.1088/0954-3899/25/9/308
[arXiv:hep-ph/9909407 [hep-ph]].

\bibitem{ALICE:2017pcy}
S.~Acharya \textit{et al.} [ALICE],
Eur. Phys. J. C \textbf{77}, no.12, 852 (2017)
doi:10.1140/epjc/s10052-017-5412-6
[arXiv:1708.01435 [hep-ex]].

\bibitem{White:1979kp}
S.~D.~M.~White,
Mon. Not. Roy. Astron. Soc. \textbf{186}, 145 (1979)

\bibitem{Glauber:1963fi}
R.~J.~Glauber,
Phys. Rev. \textbf{130}, 2529-2539 (1963)
doi:10.1103/PhysRev.130.2529

\bibitem{Wick:1950ee}
G.~C.~Wick,
Phys. Rev. \textbf{80}, 268-272 (1950)
doi:10.1103/PhysRev.80.268

\bibitem{Dall:2013}
R.~Dall \textit{et al.},
Nature Phys. 9, 341-344 (2013)
doi: 10.1038/nphys2632

\bibitem{Gotsman:2020bjc}
E.~Gotsman and E.~Levin,
Phys. Rev. D \textbf{102}, no.7, 074008 (2020)
doi:10.1103/PhysRevD.102.074008
[arXiv:2006.11793 [hep-ph]].

\bibitem{Kharzeev:2017qzs}
D.~E.~Kharzeev and E.~M.~Levin,
Phys. Rev. D \textbf{95}, no.11, 114008 (2017)
doi:10.1103/PhysRevD.95.114008
[arXiv:1702.03489 [hep-ph]].

\bibitem{Bartel}
R.~Bartel and M.~P\l oszajczak,
\textit{Universal Fluctiations, The Phenomenology of Hadronic Matter} 
(World Scientific, Singapore, 2002).

\end{thebibliography}

\end{document}